\documentclass[aps,showpacs,twocolumn,twoside,amsmath,amssymb,groupedaddress,superscriptaddress,nofootinbib,prc]{revtex4-2}
\usepackage{graphicx,here}
\usepackage{dcolumn}
\usepackage{bm}
\usepackage{hyperref}
\hypersetup{colorlinks=True,urlcolor=blue,linkcolor=blue,citecolor=blue,filecolor=black}
\begin{document}

\title{Stability of intruder-driven quadrupole and hexadecapole deformation effects 
in Xenon,  Barium, Cerium and Neodymium isotopes}

\author{K. Ziyatkhan}
\email{korlan.ziyatkhan@nu.edu.kz}

\author{R. Rodr\'{\i}guez-Guzm\'an}
\email{guzman.rodriguez@nu.edu.kz}

\affiliation{Department of Physics, School of Sciencies and Humanities, Nazarbayev 
University, 53 Kabanbay Batyr Ave., Astana 010000, Kazakhstan}

\author{L. M. Robledo}
\email{luis.robledo@uam.es}
\affiliation{%
Center for Computational Simulation, Universidad Polit\'ecnica de 
Madrid, Campus Montegancedo, 28660 Boadilla del Monte, Madrid, Spain
}%
\affiliation{Departamento  de F\'{\i}sica Te\'orica and CIAFF, 
Universidad Aut\'onoma de Madrid, 28049-Madrid, Spain}

\date{\today}

\begin{abstract} 

Two-dimensional Generator Coordinate Method calculations for the axial quadrupole 
$\beta_{2}$
and hexadecapole $\beta_{4}$ collective deformations are carried out
with the Gogny force in a series of Xe, Ba, Ce and Nd isotopes with neutron
numbers covering both magic neutron shell closures N=82 and N=126. The underlying
mean-field configurations are used to characterize the expected dynamic
behavior of the system. Two regions of strong coupling between the quadrupole
and hexadecapole degrees of freedom are found and characterized. Quantum
fluctuations soften the mean-field ground state values of $\beta_{2}$ and $\beta_{4}$
in transitional regions. The ground state correlation energy coming from
$\beta_{4}$ is comparable in size to the one coming from $\beta_{2}$, and  
both together amount to a sizable 1.5 MeV. Shape coexistence in some isotopes and its
impact in the excitation energy of the first excited state is analysed.  Finally, 
the role of a second intruder orbital in the explanation of the large deformation parameters of
some nuclei in the region is discussed. 
\end{abstract}


\maketitle

% ----------------------------------------------------------------------
%
%  S E C T I O N
%
%                                      I n t r o d u c t i o n  
%
% ----------------------------------------------------------------------  
\section{Introduction}
Understanding the role played by the
intrinsic  deformation of atomic nuclei in their collective dynamics has been 
clearly recognized for decades as one of the main challenges in nuclear structure 
physics \cite{Bohr-book,rs}. Within 
this context, quadrupole deformations which are the lowest significant multipole moments
in the multipole expansion of the intrinsic shape 
are the most extensively studied in the literature
\cite{triaxial-example-1,triaxial-example-2,triaxial-example-3,triaxial-example-4,triaxial-example-5,
triaxial-example-6,triaxial-example-7, Others-example-1,Others-example-2,Others-example-3,Others-example-4,Others-example-5,triaxal-example-8,triaxal-example-9,
triaxal-example-10,triaxal-example-11}. The impact of the quadrupole deformations of 
colliding nuclei on fusion cross sections and 
barrier distributions has also received close scrutiny 
\cite{fusion-reactions-1,fusion-reactions-2,fusion-reactions-3,
fusion-reactions-4,fusion-reactions-5} in 
nuclear 
reactions \cite{Wong-reactions,Vaz-reactions,Dasgupta-1}.

To a large extent, correlations associated with even-parity  
higher-order multipole deformation parameters, have been overshadowed 
by the large quadrupole correlations. As a result, for example, 
hexadecapole correlations have received less detailed attention in 
nuclear structure and decay studies
 \cite{triaxial-example-6,hexa-prvious-1,hexa-prvious-2,
 hexa-prvious-3,hexa-prvious-4,A.Zdeb-scission,Nomura-Lotina-b4,
 Rayner-No-Lotina-2025,NL-bb-paper}. Nevertheless, hexadecapole 
 $K^{\pi}=4^{+}$ vibrational
bands have already been identified 
\cite{exp-b4-1,exp-b4-2,exp-b4-3}, while quadrupole  and hexadecapole
deformations  have been  
inferred from   inelastic proton scattering experiments in 
$^{74}$Kr and $^{76}$Kr \cite{scattering-b2b4}. 

Previous studies have also revealed the sensitivity of fusion barrier 
distributions to the sign of the hexadecapole deformation of the target 
nucleus 
\cite{fusion-reactions-1,fusion-reactions-4,fusion-reactions-5}. 
Moreover, hexadecapole deformation has been shown to play a key role to 
improve the agreement between hydrodynamic simulations and the 
available data for collisions of $^{238}$U at the BNL Relativistic 
Heavy Ion Collider (RHIC) \cite{b4-RIHC}.

A  Hartree-Fock-Bogoliubov (HFB) survey of (axial) hexadecapole 
$\beta_{4}$ deformations in even-even nuclei, based on the Gogny 
\cite{gogny} energy density functional (EDF), was presented in 
Ref.~\cite{large-scale-b4}. The dynamical interplay between the  
quadrupole $\beta_{2}$ and hexadecapole $\beta_{4}$ deformations was 
also considered  for a selected set of Sm and Gd isotopes, in the same 
reference  within the framework of the two-dimensional (2D) Generator 
Coordinate Method (GCM) \cite{rs}. It has been shown that a better 
understanding of the nontrivial dynamical  
$(\beta_{2},\beta_{4})$-coupling requires full-fledged 2D 
configuration-mixing calculations \cite{large-scale-b4} or a careful 
definition of the relevant collective coordinates. This is certainly at 
variance with the rather weak coupling found in previous 2D-GCM 
calculations, employing the quadrupole  and octupole 
deformations as generating coordinates \cite{2DGCM-q2q3-Gogny-1,2DGCM-q2q3-Gogny-2,2DGCM-q2q3-Gogny-3,
2DGCM-q2q3-Gogny-4,2DGCM-q2q3-Gogny-5,2DGCM-q2q3-Gogny-6}.

A previous Gogny-D1S HFB+2D-GCM study 
\cite{Rayner-Robledo-b2-b4-RaThUPu} has also shown that key features, 
associated with a nontrivial beyond-mean-field $(\beta_{2}, 
\beta_{4})$-coupling, are  present in the actinide Ra, Th, U and Pu 
isotopic chains. In good agreement with the conclusions extracted in 
Ref.~\cite{b4-RIHC},  sizable static and dynamical hexadecapole 
deformation effects, associated with diamond-like shapes, have been 
obtained for ground and excited states of nuclei around  $^{238}$U. As 
expected within the polar gap model \cite{polar-gap-model}, in each of 
the studied Ra, Th, U and Pu isotopic chains, a region characterized by 
small negative hexadecapole deformation, just below the neutron magic 
number $N =184$ \cite{triaxial-example-6}, has been found to remain 
stable once zero-point $(\beta_{2}, \beta_{4})$-fluctuations are 
included within the GCM approach. Furthermore, it has been found that 
the transition from a $(\beta_{2},\beta_{4})$-coupled to an uncoupled 
regime is accompanied by an enhanced shape coexistence in the more 
neutron-rich sectors of each chain \cite{Rayner-Robledo-b2-b4-RaThUPu}.

A key  outcome from previous studies 
\cite{large-scale-b4,Rayner-Robledo-b2-b4-RaThUPu,Rayner-Robledo-b2-b4-YbHfWOs} 
is that the inclusion of hexadecapole deformation in the ground state 
dynamics leads to a correlation energy gain which is comparable to the 
quadrupole correlation energy itself. Therefore, dynamical ground state 
hexadecapole correlations might play an important role to improve our 
description of nuclear binding energies \cite{gogny-d1mstar,gogny-d1m}. 
The results of these 
\cite{large-scale-b4,Rayner-Robledo-b2-b4-RaThUPu,Rayner-Robledo-b2-b4-YbHfWOs} 
as well as previous 
\cite{2DGCM-q2q3-Gogny-1,2DGCM-q2q3-Gogny-2,2DGCM-q2q3-Gogny-3, 
2DGCM-q2q3-Gogny-4,2DGCM-q2q3-Gogny-5} 2D configuration mixing 
calculations, also point towards the slow convergence of the nuclear 
correlation energy with respect to the number of (even and/or odd) 
deformation parameters included in GCM  calculations. 

The results of previous studies 
indicate that it is timely and necessary
to deepen our understanding  of the role of hexadecapole deformations 
and their coupling to the quadrupole degree of freedom
in regions of the nuclear chart other than the previously considered Sm and Gd nuclei 
\cite{large-scale-b4},  actinides 
\cite{Rayner-Robledo-b2-b4-RaThUPu}
and large Z
rare earth elements  \cite{Rayner-Robledo-b2-b4-YbHfWOs}.

In this work, we first consider the emergence of static HFB
hexadecapole deformation effects in  the light rare earth
Xe (Z=54), Ba (Z=56), Ce (Z=58) and Nd (Z=60) nuclei. The range of neutron numbers covered 
in the selected isotopic chains, i.e., $^{112-180}$Xe, $^{116-182}$Ba, $^{118-184}$Ce
and $^{124-186}$Nd, includes the $N=82$ and $N=126$ neutron major shell 
closures and extends up to very neutron-rich sectors close 
to the corresponding two-neutron driplines. On the one hand, this allows us
to examine  the mean-field structural evolution  and onset of partial occupancy
of intruder orbits \cite{inakura-1,inakura-2}  
leading to static 
quadrupole and/or hexadecapole deformed HFB 
ground states in the considered isotopic chains. On the other hand, attention is also
paid to the emergence of square-like 
shapes just below the $N=82$ and $N=126$ magic numbers corresponding to negative hexadecapole deformations.

For the considered Xe, Ba, Ce and Nd nuclei, we have 
carried out HFB calculations 
with constrains \cite{rs} 
on the axially symmetric quadrupole $\hat{Q}_{20}$ and hexadecapole  
$\hat{Q}_{40}$ operators 
\cite{Rayner-Robledo-b2-b4-YbHfWOs}
to obtain the 
mean-field potential 
energy surfaces (MFPESs), i.e., the HFB energies $E_{HFB}(\beta_{2},\beta_{4})$
as
functions of the $\beta_{2}$ and $\beta_{4}$ deformation
parameters  (see, Secs.~\ref{Theory} and \ref{RESULTS} below). For several of 
the studied nuclei, those 
MFPESs exhibit  a soft behavior as well as a pronounced 
(quadrupole and/or hexadecapole) shape coexistence. Furthermore, the 
MFPESs also provide a  (static) glimpse on the 
$(\beta_{2},\beta_{4})$-coupling.  With this in mind, the impact of 
dynamical zero-point 
$(\beta_{2},\beta_{4})$-fluctuations has to be taken into account 
to examine, the stability of the intruder-driven
HFB quadrupole and/or hexadecapole deformation effects. Therefore, as 
a second step, the  states 
$| \varphi(\beta_{2},\beta_{4}) \rangle$, obtained 
in the $(\beta_{2},\beta_{4})$-constrained HFB calculations, are employed as a basis for the 
solution of the 2D-GCM  Griffin-Hill-Wheeler (GHW) equation \cite{rs}.
It is via the solution of the GHW equation, that 
the impact of the dynamical $(\beta_{2},\beta_{4})$-coupling 
on relevant  physical quantities is accounted for 
in the ground and excited states of 
the studied nuclei.

As in previous studies 
\cite{large-scale-b4,Rayner-Robledo-b2-b4-RaThUPu,Rayner-Robledo-b2-b4-YbHfWOs}, both 
at the HFB and 2D-GCM levels, we have resorted to the 
parametrization
D1S of the Gogny-EDF \cite{gogny,review-RRR-2019}. Let us stress, however, that
rather similar results have been obtained in calculations 
with the parametrizations D1M$^{*}$ \cite{gogny-d1mstar} and D1M 
\cite{gogny-d1m}. The  similarity in the results corroborates the robustness of the observed trends
with respect to the underlying Gogny-EDF.

The paper is organized as follows. The HFB+2D-GCM framework 
\cite{2DGCM-q2q3-Gogny-1,2DGCM-q2q3-Gogny-2, 
2DGCM-q2q3-Gogny-3,2DGCM-q2q3-Gogny-4,2DGCM-q2q3-Gogny-5,large-scale-b4, 
Rayner-Robledo-b2-b4-RaThUPu,Rayner-Robledo-b2-b4-YbHfWOs} is briefly 
outlined in Sec.~\ref{Theory}. 
The results of Gogny-D1S  mean-field HFB and beyond-mean-field 2D-GCM calculations are discussed 
in Sec.~\ref{RESULTS}. First, in Sec.~\ref{static-HFB-TeXeBaCeNd}, attention is 
paid to  static quadrupole and hexadecapole deformations, the 
role of intruder orbits in the emergence of those deformations as well 
as to the structural evolution of the   MFPESs. The $(\beta_{2},\beta_{4})$-GCM
results obtained for  Xe, Ba, Ce and Nd nuclei are discussed in
Sec.~\ref{dynamical-GCM-TeXeBaCeNd}. Here, attention is paid 
to the impact of zero-point 
$(\beta_{2},\beta_{4})$-fluctuations on the stability of the 
HFB predictions via the analysis of 
ground state collective wave 
functions, dynamical quadrupole and hexadecapole  
deformations, correlation energies as well 
as the comparison with different types of one-dimensional (1D) GCM  
calculations. The role of the dynamical $(\beta_{2},\beta_{4})$-coupling 
in the 2D-GCM  first excited states of the considered nuclei is 
also examined in this section. Finally, 
Sec.~\ref{conclusions} is devoted to the concluding remarks.

% -----------------------------------------------------------------------
%
%                              T h e o r e t i c a l   F r a m e w o r k 
%
% -----------------------------------------------------------------------
\section{Theoretical framework}
\label{Theory}
% -----------------------------------------------------------------------
The HFB+2D-GCM scheme employed 
in this study is briefly outlined 
in this section. For more details, the reader is referred, for example, to
Ref.~\cite{Rayner-Robledo-b2-b4-RaThUPu} and references therein.
In order to obtain the $(\beta_{2},\beta_{4})-$MFPESs as well as a set of 
mean-field states $| \varphi(\beta_{2},\beta_{4}) \rangle$ for the 2D-GCM calculation, the 
Gogny-D1S \cite{gogny,review-RRR-2019} HFB equation has been
solved \cite{gradient} with constrains on the mean values of   
the axially symmetric quadrupole $\hat{Q}_{20}$ and hexadecapole  
$\hat{Q}_{40}$ operators \cite{Rayner-Robledo-b2-b4-YbHfWOs}.
The mean values $Q_{\lambda 0}$ ($\lambda$=2,4)
of those constraining operators in the intrinsic
HFB states $| \varphi \rangle$, have been 
written 
in terms of the standard deformation parameters
$\beta_{\lambda}$ \cite{large-scale-b4,Rayner-Robledo-b2-b4-RaThUPu}.

In the constrained Gogny-HFB calculations, axial symmetry has been kept 
as a selfconsistent symmetry \cite{rs}. We have obtained intrinsic HFB 
states $| \varphi(\beta_{2},\beta_{4}) \rangle$ and energies 
$E_{HFB}(\beta_{2},\beta_{4})$ in a large $(\beta_{2},\beta_{4})$-mesh, 
with $\beta_{2} \in [-0.80,0.92]$, $\beta_{4} \in [-0.80,0.92]$ and 
step sizes $\delta \beta_{2}= \delta \beta_{4}= 0.02$. We have 
considered such a large number of HFB 
$(\beta_{2},\beta_{4})$-configurations to account for all types of 
minima in the MFPESs as well as shape coexistence and shape 
transitions.

The HFB quasiparticle operators \cite{rs} have been expanded in an 
axially symmetric harmonic oscillator (HO) basis consisting of 
$N_{shell}$=17 shells. Note, that our calculations extend to very 
neutron-rich sectors of the studied isotopic chains. Therefore, a large 
number of HO shells is employed to guarantee the convergence of the 
potential energy curves $E_{HFB}(\beta_{2},\beta_{4})$ associated with 
different types of $(\beta_{2},\beta_{4})$-configurations. All the HFB 
states $| \varphi(\beta_{2},\beta_{4}) \rangle$ have been computed with 
the same oscillator lengths $b_{\perp}=b_{z}= b_{0}=1.01 A^{1/6}$ to 
facilitate the evaluation of  2D-GCM kernels \cite{EWT-1,EWT-2}.

The 
HFB states $| \varphi(\vec{\beta}) \rangle$, with $\vec{\beta}= (\beta_{2},\beta_{4})$, are employed as a basis 
in the  2D-GCM ansatz
\begin{equation} \label{GCM-WF}
| {\Psi}_{2D-GCM}^{\sigma} \rangle = \int d\vec{\beta} f^{\sigma} (\vec{\beta}) | {\varphi} (\vec{\beta}) \rangle
\end{equation}
where, the index $\sigma$ numbers the
ground ($\sigma =1$) and excited ($\sigma =2, \cdots$)
states. The variation of the 2D-GCM energy  with respect
to the amplitudes $f^{\sigma} (\vec{\beta})$ leads to the 
GHW equation 
\begin{equation}
{\cal{H}} f^{\sigma} = E^{\sigma} {\cal{N}} f^{\sigma}
\end{equation}
where ${\cal{H}}(\vec{\beta}_{1},\vec{\beta}_{2})$ and ${\cal{N}}(\vec{\beta}_{1},\vec{\beta}_{2})$ represent the Hamiltonian and 
norm kernels \cite{rs}. In the computation of the Gogny Hamiltonian  kernel, we
have employed the mixed-density prescription  \cite{denpres,projden}
as well as first-order corrections in the  mean values of the proton and neutron 
numbers of the correlated wave function \cite{2DGCM-q2q3-Gogny-1,2DGCM-q2q3-Gogny-2,
2DGCM-q2q3-Gogny-3,2DGCM-q2q3-Gogny-4,2DGCM-q2q3-Gogny-5,large-scale-b4,
Rayner-Robledo-b2-b4-RaThUPu,Rayner-Robledo-b2-b4-YbHfWOs}. 

At the 2D-GCM level, we have performed calculations with the same mesh 
$\beta_{2} \in [-0.80,0.92]$, $\beta_{4} \in [-0.80,0.92]$. However, in 
this case we have resorted to the step sizes $\delta \beta_{2}= \delta 
\beta_{4}= 0.04$ to alleviate the computational effort. We are then 
left with 44 grid points for each of the $\beta_{2}$ and $\beta_{4}$ 
generating coordinates. There is some amount of linear dependency in 
the set of 1936 basis $(\beta_{2},\beta_{4})$-configurations. Such a 
linear dependency is treated using the standard procedure, i.e., only 
states in the natural GCM basis with norm eigenvalues greater that 
$10^{-4}$ are selected \cite{rs}. We have checked that the selected 
$(\beta_{2},\beta_{4})$-mesh is enough to account for the properties of 
the 2D-GCM ground and first excited states in which we concentrate for 
all the studied nuclei.

In order to have access to a  probabilistic interpretation in terms of
the collective variables $\beta_{2}$ and $\beta_{4}$, we have 
introduced the collective wave functions 
\begin{equation} \label{cll-wfs-HW} 
G^{\sigma} (\vec{\beta}_{1}) =   \int d\vec{\beta}_{2} ~
{\cal{N}}^{\frac{1}{2}}(\vec{\beta}_{1},\vec{\beta}_{2}) 
f^{\sigma} (\vec{\beta}_{2})
\end{equation}
written in terms of the amplitudes $f^{\sigma}$  and 
the operational square root 
of the norm kernel ${\cal{N}}^{\frac{1}{2}}$ \cite{rs}.  
With the collective wave functions  
at hand, dynamical quadrupole 
$\beta_{2,2D-GCM}^{\sigma}$
and hexadecapole 
$\beta_{4,2D-GCM}^{\sigma}$
deformations  have been obtained as already 
discussed in Ref.~\cite{Rayner-Robledo-b2-b4-RaThUPu}.
Other relevant  physical quantities have been computed according to the general 
expressions employed in previous works \cite{2DGCM-q2q3-Gogny-1,2DGCM-q2q3-Gogny-2,
2DGCM-q2q3-Gogny-3,2DGCM-q2q3-Gogny-4,2DGCM-q2q3-Gogny-5,large-scale-b4,
Rayner-Robledo-b2-b4-RaThUPu,Rayner-Robledo-b2-b4-YbHfWOs}.

We are aware that octupolarity can play a role 
around some of the proton $Z$ and/or neutron $N$ numbers
covered in the studied isotopic chains \cite{Naza-Q3}.
However, 
previous studies 
in different regions of the nuclear chart 
\cite{2DGCM-q2q3-Gogny-1,2DGCM-q2q3-Gogny-2,2DGCM-q2q3-Gogny-3,
2DGCM-q2q3-Gogny-4,2DGCM-q2q3-Gogny-5}
have shown that the quadrupole-octupole coupling is rather weak. 
As a result of this weak 
coupling, essential features associated with the octupole 
dynamics can be accounted for using 1D-GCM 
calculations, with the octupole deformation as the only generating 
coordinate. In particular, this has already been shown, via Gogny
quadrupole-octupole
2D-GCM calculations
for Xe, Ba, Ce and Nd isotopes with neutron numbers
$54 \le N \le 96$ \cite{2DGCM-q2q3-Gogny-6}. 

The weak quadrupole-octupole coupling arises from the different parity 
quantum numbers of the multipole moments involved. For the same reason, 
an even weaker hexadecapole-octupole coupling can be anticipated. With 
this in mind, in this study both HFB and 2D-GCM calculations will be 
carried out only in terms of the  $\beta_{2}$ and $\beta_{4}$ 
deformation parameters.  Note also, that the consideration of 
triaxiality and/or symmetry restoration will make the calculations, at 
the present stage, prohibitive expensive. Work along these lines is in 
progress and will be reported in future publications.

% ----------------------------------------------------------------------
%
%                      D i s c u s s i o n   o f   t h e   r e s u l t s 
%
%
% ----------------------------------------------------------------------

\section{Discussion of the results}
\label{RESULTS}
The results of the calculations for 
$^{112-180}$Xe, $^{116-182}$Ba, $^{118-184}$Ce
and $^{124-186}$Nd are discussed in
this section. First, HFB results will be presented
in Sec.~\ref{static-HFB-TeXeBaCeNd}, while 2D-GCM results will be discussed in
Sec.~\ref{dynamical-GCM-TeXeBaCeNd}. 

% ----------------------------------------------------------------------
%                                                                Figure 
% ---------------------------------------------------------------------- 
\begin{figure}
\includegraphics[width=0.48\textwidth]{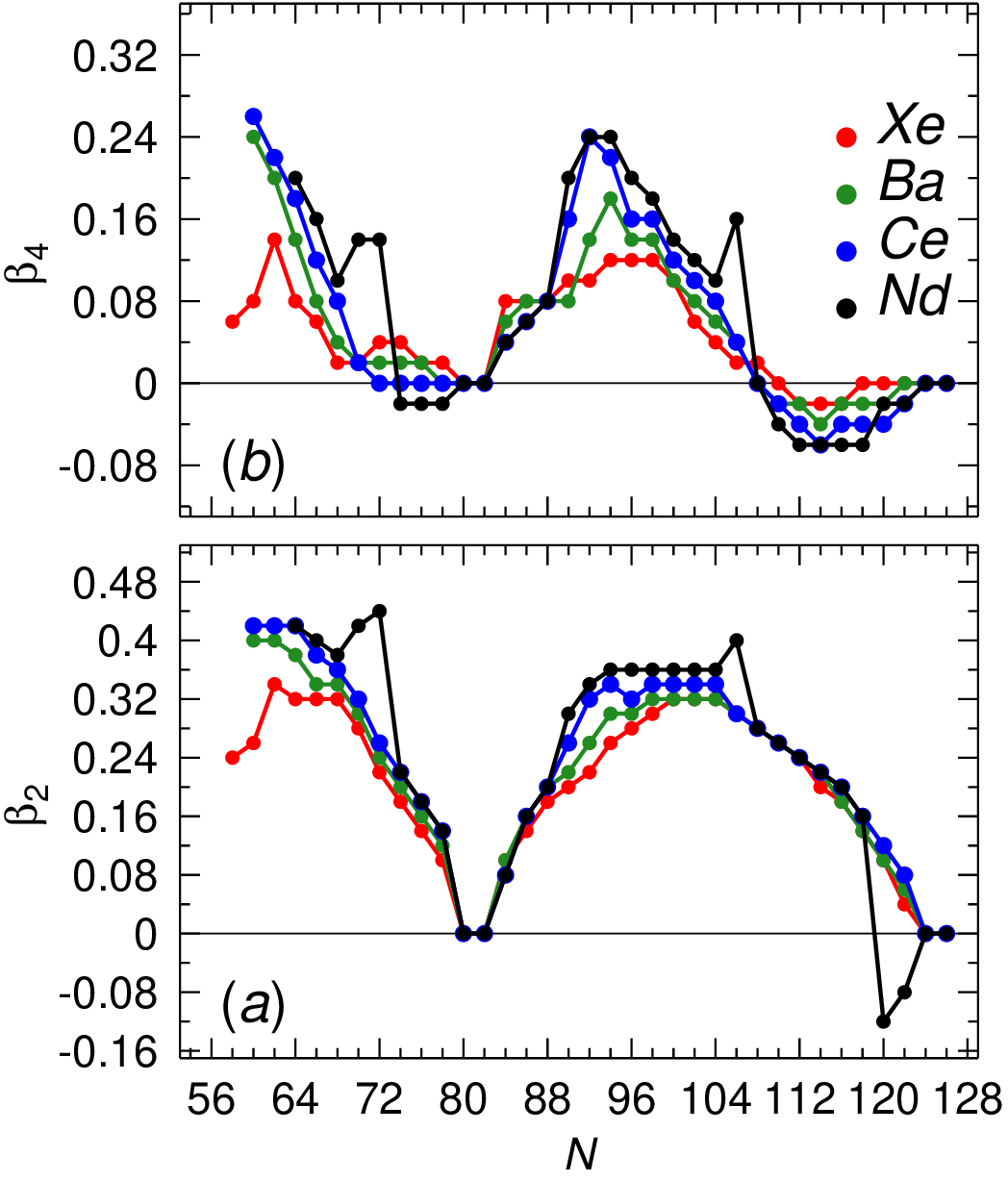}
\caption{(Color online) The HFB ground state quadrupole 
$\beta_{2}$
[panel (a)]
and hexadecapole  $\beta_{4}$ [panel (b)] deformations 
obtained for the nuclei
 $^{112-180}$Xe, $^{116-182}$Ba, $^{118-184}$Ce
 and $^{124-186}$Nd are plotted as functions of the neutron number.
 Results have been obtained with the Gogny-D1S EDF.
}
\label{HFB-b2b4-gs-def} 
\end{figure}

% ----------------------------------------------------------------------
%                                                   Subsect HFB results
% ---------------------------------------------------------------------- 

\subsection{The HFB approach for Xe, Ba, Ce and Nd nuclei}
\label{static-HFB-TeXeBaCeNd}

As already mentioned, $(\beta_{2},\beta_{4})$-constrained Gogny-D1S HFB calculations have been performed 
to obtain the corresponding MFPESs for the studied 
Xe, Ba, Ce and Nd isotopes. The 
quadrupole $\beta_{2}$ and hexadecapole $\beta_{4}$ deformation parameters corresponding to 
the absolute minima of those MFPESs are depicted in panels (a)
and (b) of Fig.~\ref{HFB-b2b4-gs-def}, as functions 
of the neutron number $N$.

In panel (a), the quadrupole deformations for Xe isotopes initially 
increase, reaching $\beta_{2}=0.34$ in $^{116}$Xe (N=62). For larger neutron numbers those 
deformations decrease, reaching $\beta_{2}=0$ in 
$^{134,136}$Xe in the vicinity of the neutron magic number N=82 .  In the case of Ba and Ce isotopes, the quadrupole deformations 
decrease smoothly, reaching $\beta_{2}=0$ in the vicinity of the neutron magic number N=82, that is  $^{136,138}$Ba and $^{138,140}$Ce. A similar overall 
decrease towards  
$\beta_{2}=0$ ($^{140,142}$Nd)
is observed in the Nd isotopic chain, although a local and abrupt increase in the 
$\beta_{2}$ values appears for the $N=70,72$ isotopes. 

% ----------------------------------------------------------------------
%                                                                Figure 
% ---------------------------------------------------------------------- 

\begin{figure}
	\includegraphics[width=1.0\columnwidth]{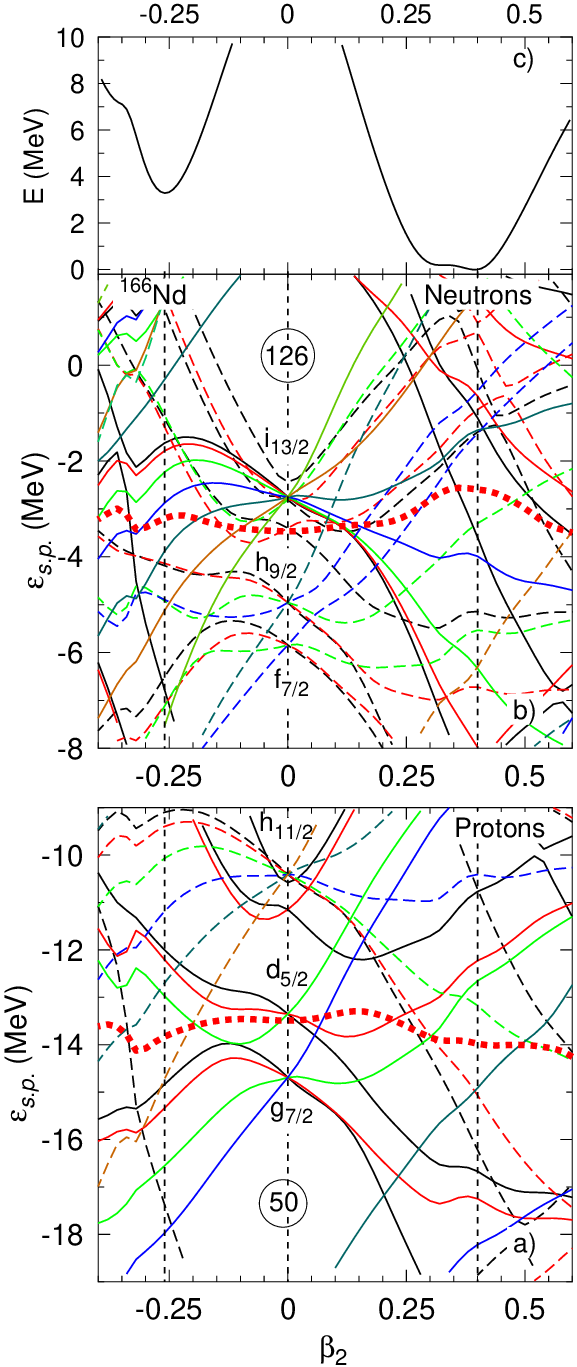}
    \caption{(Color online) Single particle energies are plotted for both protons [panel (a)]
    and neutrons [panel (b)] as a function of $\beta_{2}$ in the nucleus
    $^{166}$Nd. Full (dashed) lines are for positive (negative) parity
    orbitals. The $K$ quantum number of each orbital is characterized by 
    a color code: black $K=1/2$, red $K=3/2$, green $K=5/2$, blue $K=7/2$, dark grey $K=9/2$, brown $K=11/2$, and forest green $K=13/2$.
     The Fermi level is depicted as a thick (red) dotted line. In panel (c) the
    HFB energy is plotted, as a function of $\beta_{2}$, to identify the
    position of the prolate and oblate minima as well as the spherical
    configuration.}
    \label{166Ndspe}
\end{figure}

After passing the neutron major shell closure  
N=82, the quadrupole deformation increases, reaching its largest values
($\beta_{2}=0.32-0.36$) around $N=102$, with a  kink 
($\beta_{2}=0.40$) for $^{166}$Nd (N=106). For larger neutron numbers, the 
$\beta_{2}$ parameter decreases, reaching $\beta_{2}=0$ in the vicinity of 
neutron magic number N=126 for the nuclei
$^{178,180}$Xe, $^{180,182}$Ba, $^{182,184}$Ce and $^{184,186}$Nd. Note, that
oblate ($\beta_{2}=-0.12$ and $-0.08$)
ground states are obtained for $^{180,182}$Nd just before reaching N=126.

In panel (b), the hexadecapole deformations exhibit a trend 
similar to that of the quadrupole deformations. This already suggests a
certain 
$(\beta_{2},\beta_{4})$-interrelation at the HFB level \cite{inakura-2}. For Xe isotopes, the 
hexadecapole deformations initially 
increase, reaching $\beta_{4}=0.14$ 
in $^{116}$Xe. With increasing neutron number, 
$\beta_{4}$ then shows an overall decrease, reaching $\beta_{4}=0$ around 
N=82 in $^{134,136}$Xe. In the case of Ba and Ce isotopes, diamond-like ($\beta_{4} > 0$)
ground 
states are predicted, with decreasing $\beta_{4}$ deformations for increasing neutron number,
that reach $\beta_{4}=0$ in $^{134-138}$Ba and $^{130-140}$Ce (around N=82).
In the Nd isotopic chain, a local increase in the 
$\beta_{4}$ values appears for the $N=70,72$ isotopes in correspondence with 
the abrupt increase in the HFB  ground state $\beta_{2}$ deformation in these two
isotopes. This sudden increase in $\beta_{4}$ is  followed by 
negative hexadecapole deformations 
($\beta_{4}=-0.02$) in $^{134-138}$Nd, just below the neutron magic number 
N=82  \cite{polar-gap-model}, and by $\beta_{4}=0$ in $^{140,142}$Nd.

The hexadecapole deformations 
increase beyond  N=82, reaching their largest values 
$\beta_{4}=0.12, 0.18, 0.24$ 
and $0.24$ in $^{148-152}$Xe, $^{150}$Ba, $^{150}$Ce and
$^{152,154}$Nd, respectively. For  larger neutron numbers, those deformations 
decrease, exhibiting a kink ($\beta_{4}=0.16$) in
$^{166}$Nd. As the neutron number approaches  
N=126, a region of negative 
($-0.06 \le \beta_{4} \le -0.02$)
hexadecapole deformations 
\cite{polar-gap-model}
emerges in all the studied isotopic chains. In particular, at
the HFB level, the nuclei in this $\beta_{4} < 0$ region 
are $^{166-170}$Xe, $^{166-176}$Ba, $^{168-180}$Ce and 
$^{170-182}$Nd. 

From the previous analysis one can conclude that large and positive 
hexadecapole deformations appear associated to large  prolate quadrupole
deformations whereas small or even negative values of $\beta_{4}$ are to
be associated to small quadrupole deformations of prolate or oblate character
specially for neutron numbers below magic values.

% ----------------------------------------------------------------------
%                                                                Figure 
% ---------------------------------------------------------------------- 

\begin{figure}
	\includegraphics[width=1.0\columnwidth]{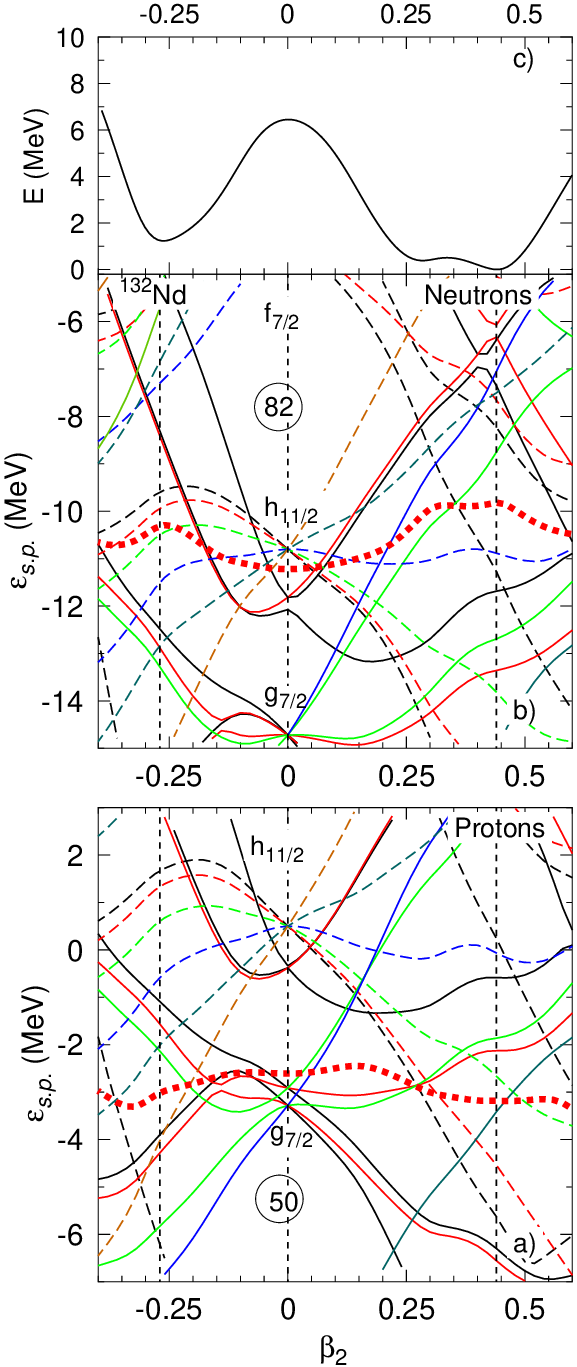}
    \caption{(Color online) The same as in Fig.~\ref{166Ndspe}, but 
    for the nucleus $^{132}$Nd.}
    \label{132Ndspe}
\end{figure}

In order to better understand the peculiar behavior of $^{166}$Nd regarding
its ground state deformation values we have plotted the evolution of the
single particle energies for both protons [panel (a)] and neutrons 
[panel (b)]
as a function of 
$\beta_{2}$ in Fig.~\ref{166Ndspe}. In panel (c) of the figure, we have also depicted the
HFB energy, as a function of $\beta_{2}$, to identify the
position of the prolate and oblate minima as well as the spherical
configuration. As discussed 
in Refs. \cite{inakura-1,inakura-2}, the existence of large ground
 state quadrupole deformation can be associated
with a ``second intruder" orbital with a large extension into the 
$z$ axis 
\footnote{Orbitals with predominant Nilsson quantum numbers corresponding to large $n_{z}$ values.} 
plunging into the Fermi sea. 
This is indeed the case in $^{166}$Nd where 
deformation is partly driven by the occupancy of the $K=1/2$, 3/2 and 5/2 
components of the ``first intruder" orbital $i_{13/2}$  followed by
the occupancy of an [660]1/2 orbital coming for the ``second intruder" $g_{7/2}$
spherical orbit. This orbital replaces a $K=5/2^{-}$ orbital from the spherical $h_{9/2}$ at precisely $\beta_{2}=0.4$ and a 
bit later  it crosses  a $K=7/2^{+}$ orbital from the spherical $i_{13/2}$.
The contributions of the second intruder orbital to the quadrupole $Q_{20}$ and 
hexadecapole $Q_{40}$ multipole moments at $\beta_{2}=0.4$ are 31.2 fm$^{2}$ and 1258 fm$^{4}$, respectively.
These numbers have to be compared to the contributions from the $K=7/2^{+}$ 
component of the first intruder 11.16 fm$^{2}$ and -316 fm$^{4}$, respectively.
On the other hand, the values for the $K=5/2^{-}$ orbital 
from the spherical $h_{9/2}$ are -0.45 fm$^{2}$ and -316 fm$^{4}$.
Clearly, by replacing the $K=5/2^{-}$ and $K=7/2^{+}$ orbitals by the $K=1/2$ from 
the second intruder one favors larger quadrupole and hexadecapole deformations. One has to 
take into account that the previous 
discussion, based on the diagonal matrix elements of $Q_{20}$ and  $Q_{40}$, does not
take into account the polarization effects induced by the second intruder in 
the other orbitals. In any case, our Gogny-D1S results seem to confirm the 
arguments of Refs. \cite{inakura-1,inakura-2} concerning the role of 
the second intruder in the very large quadrupole deformation observed in
some isotopes. 

Another nucleus of interest is $^{132}$Nd which is the one with the largest
$\beta_{2}$ and $\beta_{4}$ values for the ground state minimum. In Fig.~\ref{132Ndspe}, the
single particle energies for both protons [panel (a)] and neutrons 
[panel (b)]are plotted as
a function of $\beta_{2}$. In panel (c) of the figure, we have also depicted the
HFB energy, as a function of $\beta_{2}$, to identify the
position of the prolate and oblate minima as well as the spherical
configuration. Contrary to the $^{166}$Nd case, the intruder
orbital for neutrons is the negative parity $h_{11/2}$ which is partially occupied in
this isotope with N=72. The second intruder orbital, which is above the N=82
shell closure for spherical configurations, is the $f_{7/2}$ orbital. 
When
prolate deformation sets in, the energy of the $K=1/2$ component of this orbital 
decreases quickly and hits the Fermi level at a deformation of $\beta_{2}=0.35$ 
becoming fully occupied afterwards and giving rise to the ground state
minimum located at $\beta_{2}=0.44$. As in the case of $^{166}$Nd  the
occupancy of the second intruder comes together with the emptying of
the $K=7/2^{-}$ from the $h_{11/2}$ and, to a lesser extent, with the
emptying of the $K=5/2^{+}$ from the $g_{7/2}$ spherical orbital.

%######################################################################
% figure 
%######################################################################
\begin{figure*}
\includegraphics[width=1.\textwidth]{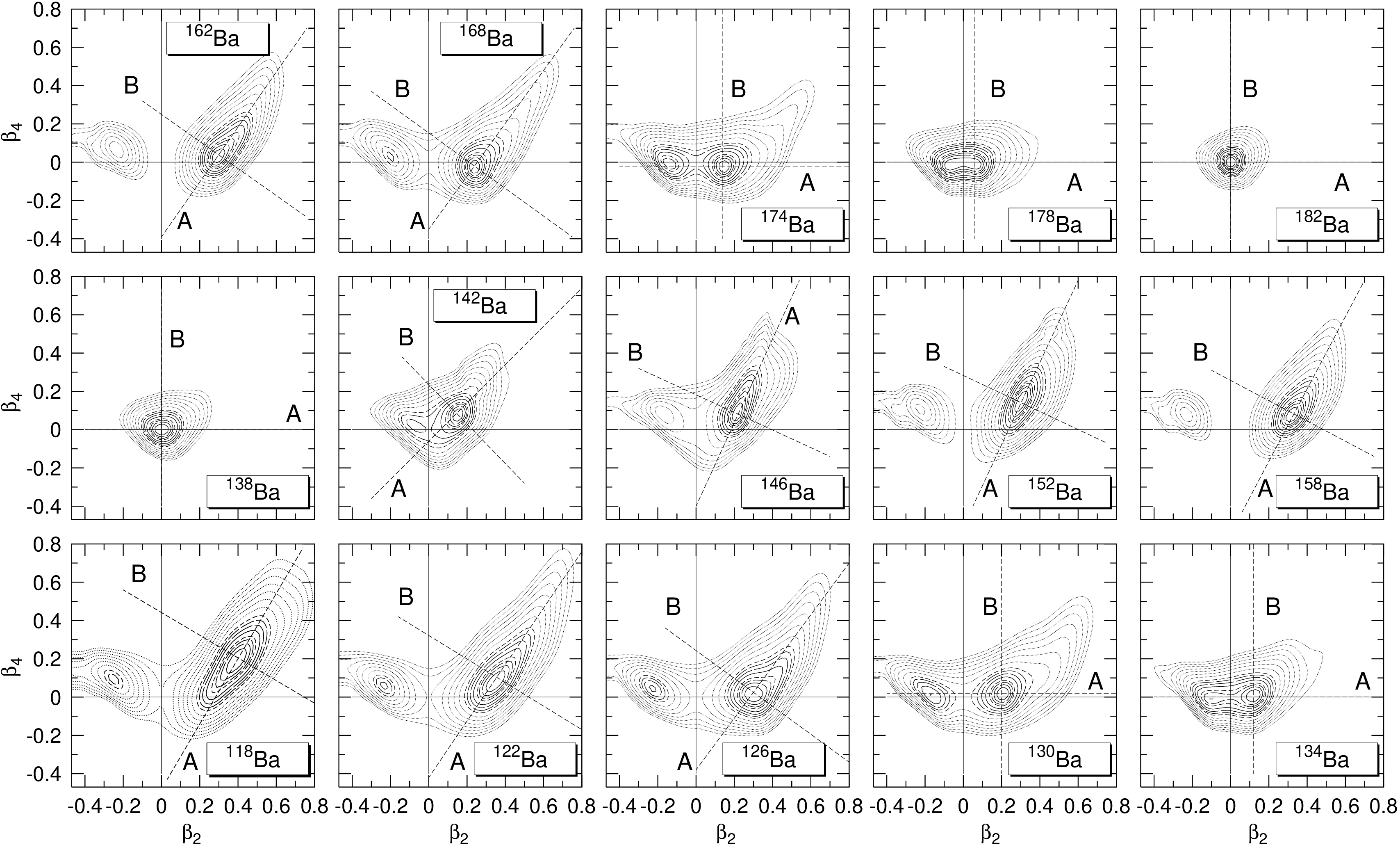}
\caption{MFPESs computed with the Gogny-D1S EDF for a selected
set of Ba isotopes. Contour lines extend from 0.25 MeV up 
to 1 MeV above the ground state energy in steps of 0.25 MeV in the 
ascending sequence full, long-dashed, medium-dashed and short-dashed. 
The next contours following the same sequence correspond to
energies from 1.5 MeV up to 3 MeV above the 
ground state in steps of 0.5 MeV. From there on, 
dotted contour lines are drawn in steps of 1 MeV. For each nucleus, the two 
perpendicular dotted lines A and B are drawn along the principal axes 
of the parabola that approximates the HFB energy around the absolute 
minimum of the MFPES. A vertical full line is drawn to signal the $\beta_{2}=0$
line whereas a full horizontal line is drawn to signal the $\beta_{4}=0$ line.  
For more details, see the main text.
}
\label{MFPES-examplesBa} 
\end{figure*}

The analysis of the structural evolution of the MFPESs provides a 
useful (static) glimpse  on the coupling between the quadrupole and 
hexadecapole degrees of freedom in the considered nuclei. In 
Fig.~\ref{MFPES-examplesBa}, we have plotted such MFPESs for a selected 
set of Ba isotopes, taken as representative examples. A similar 
analysis has been carried out for Xe, Ce and Nd nuclei.

From Fig.~\ref{MFPES-examplesBa}, one realizes that for $^{118,122,126}$Ba 
the bottom of the energy valley around the absolute 
minimum of the MFPES aligns along the direction represented by 
the dotted line A, which is tilted with respect to the 
$\beta_{2}$-axis. The dotted line B, tilted 
with respect to the $\beta_{4}$-axis, represents 
the direction perpendicular to A. We have found  that the HFB energies 
along the directions A and B
exhibit a parabolic behavior as functions 
of $\beta_{2}$ \cite{Rayner-Robledo-b2-b4-RaThUPu}. Moreover, the $\beta_{4}$ deformations 
depend linearly on the quadrupole deformation along those 
directions A and B.
For 
example, for $^{122}$Ba, we have 
obtained  $\beta_{4,A}= 1.48 \beta_{2} - 0.42$ and 
$\beta_{4,B}= -0.68 \beta_{2} + 0.31$, using 
configurations around the absolute minimum of the MFPES. 
A similar procedure has been followed to fit the lines 
A and B in other nuclei   
\cite{Rayner-Robledo-b2-b4-RaThUPu,Rayner-Robledo-b2-b4-YbHfWOs}.
We have found that for Ba isotopes
with $60 \le N \le 72$, the quadrupole and hexadecapole degrees 
are coupled. In such nuclei 1D-GCM calculations using the 
$\beta_{2}$  or $\beta_{4}$
deformations as single generating 
coordinates (i.e., $\beta_{2}$-GCM or $\beta_{4}$-GCM)
essentially explore the same configurations around 
the absolute minimum of the MFPES corresponding to those in the direction A.
Therefore, to fully take into account the  
$(\beta_{2},\beta_{4})$-coupling a full-fledged 2D-GCM calculation with $(\beta_{2},\beta_{4})$ as
generating coordinates must be carried out 
\cite{large-scale-b4,Rayner-Robledo-b2-b4-RaThUPu,Rayner-Robledo-b2-b4-YbHfWOs}. 
A cheaper alternative to this ambitious program is to carry out 1D-GCM calculations
along the directions A and B as they are truly independent variables, at 
least around the ground state minimum. The results of such 1D-GCM calculations
will be discussed later.

The nuclei $^{130,134,138}$Ba fall within a 
$(\beta_{2},\beta_{4})$-decoupled regime, corresponding to 
$74 \le N \le 82$, in which the lines A and B run parallel to the 
$\beta_{2}$ and $\beta_{4}$ axes, respectively, and therefore $\beta_{2}$ and $\beta_{4}$ can be used as
independent variables around the ground state minimum. In contrast, isotopes such as 
$^{142,146,152,158,162,168}$Ba, which are representatives of the region 
$84 \le N \le 112$, exhibit directions A and B   tilted with respect to the 
$\beta_{2}$ and $\beta_{4}$ axes, indicating again a 
$(\beta_{2},\beta_{4})$-coupled regime. Finally, in the heaviest Ba isotopes with 
$114 \le N \le 126$, the quadrupole and hexadecapole degrees of freedom are decoupled.

From our analysis of all the Gogny-D1S MFPESs considered in this paper we 
observe the same pattern as in the Ba isotopic chain: two regions where 
the quadrupole and hexadecapole degrees of freedom are interwoven and
another two where both degrees of freedom are decoupled. The first transition  
from a coupled to a decoupled regime 
takes place for $N=70, 74, 76$ and $78$ in 
Xe, Ba, Ce and Nd isotopes. On the other hand, in 
Xe, Ba and  Ce nuclei the second transition occurs for $N=114$ while 
in the Nd chain it occurs for $N=118$.

\begin{figure}
\includegraphics[width=0.48\textwidth]{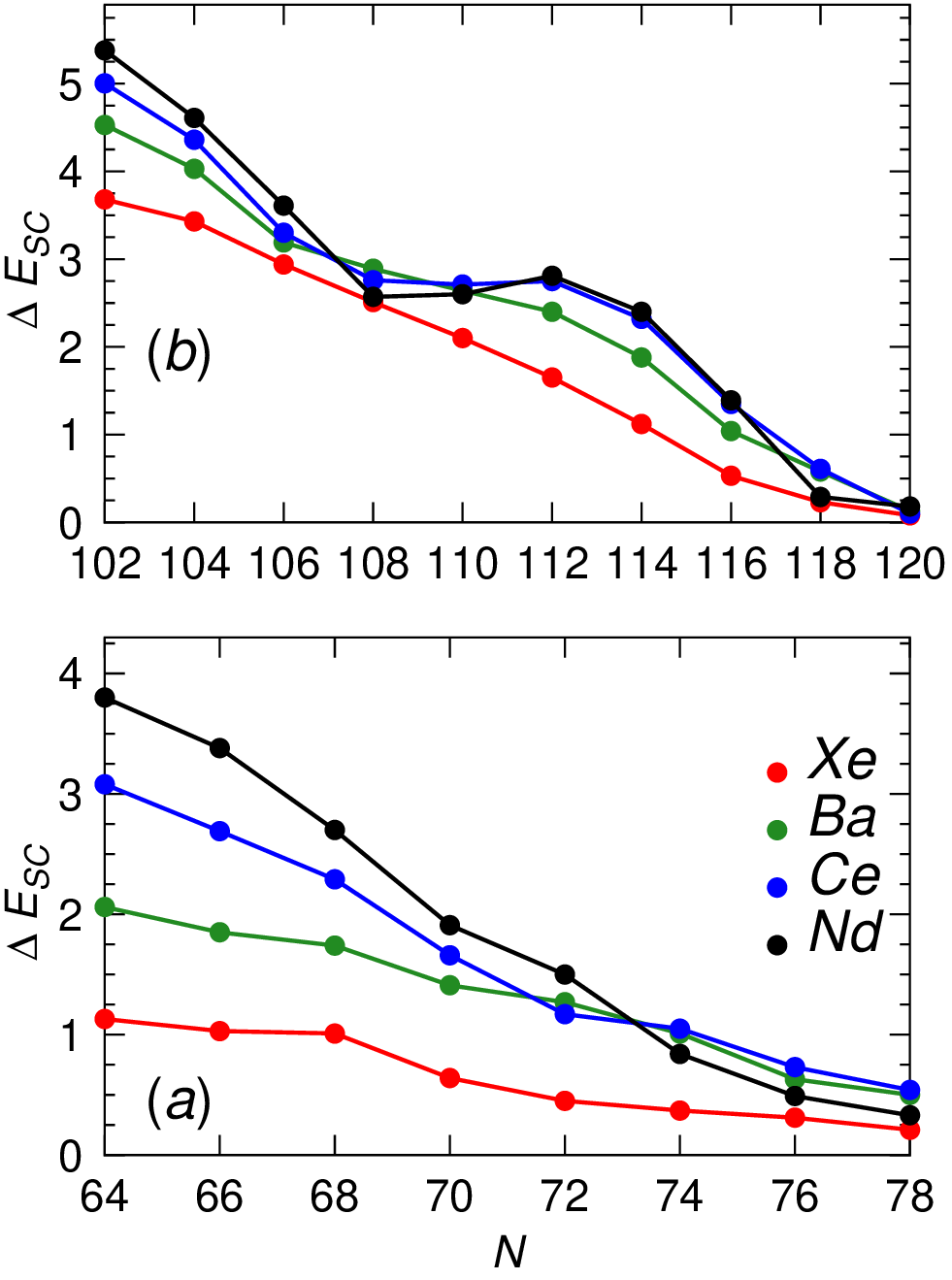}
\caption{(Color online) The difference $\Delta E_{SC}$
between the HFB energies corresponding to the absolute 
and secondary minima of the MFPESs are plotted 
for Xe, Ba, Ce and Nd isotopes with neutron numbers
$64 \le N \le 78$ [panel (a)] and $102 \le N \le 120$ [panel (b)].
Results have been obtained with the Gogny-D1S EDF.
}
\label{SCoexistence} 
\end{figure}

In panels (a) and (b) of Fig.~\ref{SCoexistence}, we show the difference 
$\Delta E_{SC}$
between the HFB energies associated with the absolute and secondary minima of the MFPESs
for Xe, Ba, Ce and Nd isotopes with neutron numbers 
 $64 \le N \le 78$  and $102 \le N \le 120$, respectively. In all the
nuclei discussed in this paper, exception made of $^{{180}}$Nd the 
absolute minimum is always prolate (or spherical) and the secondary 
one, oblate (or spherical).  The pronounced reduction in $\Delta 
E_{SC}$, as one moves towards the N=82 and 126 shell closures, 
indicates that the transitions from a $(\beta_{2},\beta_{4})$-coupled 
to an uncoupled regime are accompanied by an enhanced competition 
between configurations characterized by different intrinsic quadrupole 
and/or hexadecapole deformations, i.e., shape coexistence. Furthermore, 
as shown in panel (b), the $\Delta E_{SC}$ values for the Nd and Ce 
isotopic chains exhibit a minimum around $N=108$. This minimum becomes 
less prominent  for Ba and is absent in the Xe isotopes. As will be 
shown later on in the paper (see, Sec.~\ref{sec-exc-states}) 
fingerprints of the subtle balance between competing configurations are 
found in the energies of the 2D-GCM first excited states for the 
considered nuclei. Furthermore, it should also be noted that, for 
several of the studied nuclei, the MFPESs display a soft behavior along 
the quadrupole and/or hexadecapole directions.

The features already discussed in this section -namely, quadrupole 
and/or hexadecapole shape transitions, transitions between different 
$(\beta_{2},\beta_{4})$-coupling regimes,  enhanced shape coexistence 
and softness in the MFPESs- clearly indicate the need to account for 
beyond-mean-field $(\beta_{2},\beta_{4})$-fluctuations within a 
dynamical 2D-GCM framework. It should also be emphasized that, in such 
a dynamical configuration-mixing approach, not only the topography of 
the MFPES but also the underlying inertial properties associated with 
the generating $\beta_{2}$ and $\beta_{4}$ collective coordinates play 
a  role \cite{rs}. Such  2D-GCM calculations are also required to 
assess the dynamical stability of $\beta_{4}<0$ regions in the studied 
isotopic chains as the $N=82$ and $126$ magic numbers are approached. 
Note that, exception made of the previous studies 
\cite{Rayner-Robledo-b2-b4-RaThUPu,Rayner-Robledo-b2-b4-YbHfWOs}, a 
detailed dynamical treatment of negative hexadecapole deformations has 
not yet been reported in the literature.
 
%######################################################################
% figure 
%######################################################################

% ----------------------------------------------------------------------
%                                Subsection
%
% ----------------------------------------------------------------------

\subsection{The 2D-GCM approach for Te, Xe, Ba, Ce and Nd nuclei}
\label{dynamical-GCM-TeXeBaCeNd} 

\subsubsection{Ground states}

\begin{figure*}
\includegraphics[width=1.\textwidth]{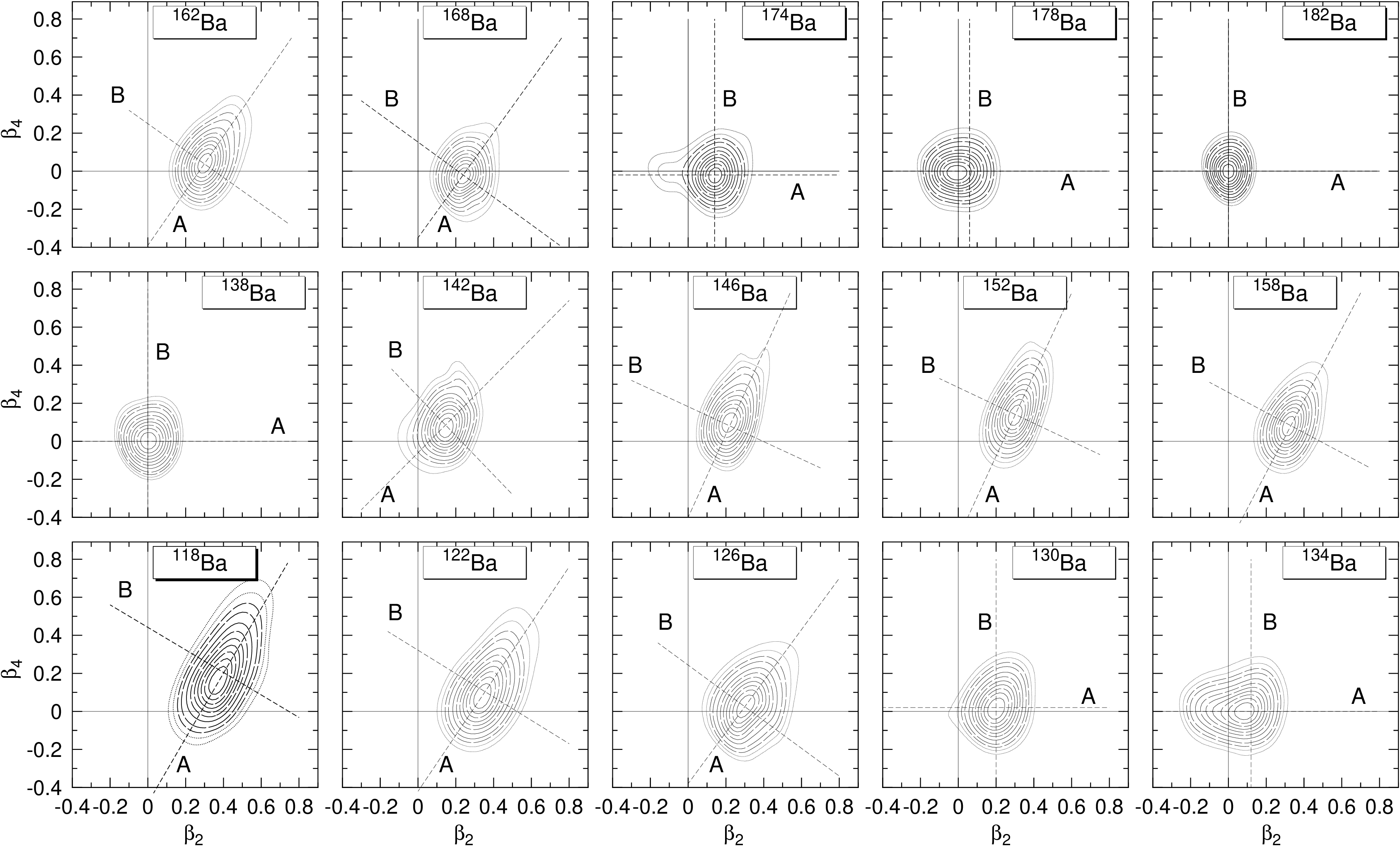}
\caption{Collective wave functions Eq.(\ref{cll-wfs-HW}) corresponding 
to the ground states for a selected
set of Ba isotopes (see, Fig.~\ref{MFPES-examplesBa}). The succession of solid, long 
dashed and short dashed contour 
lines starts at 90$\%$ of the maximum value up to 10$\%$ of it. The two 
dotted-line contours correspond to the tail of the amplitude (5$\%$ and 
1$\%$ of the maximum value).  For each nucleus, the two perpendicular 
dotted lines A and B are drawn along the principal axes of the parabola 
that approximates the HFB energy around the absolute minimum of the 
MFPES and are the same as in Fig.\ref{MFPES-examplesBa}. A 
vertical full line is drawn to signal the $\beta_{2}=0$ line whereas a 
full horizontal line is drawn to signal the $\beta_{4}=0$ line. Results 
are obtained with the Gogny-D1S EDF. For more details, see the 
main text.
}
\label{GSCWF-examplesBa} 
\end{figure*}

Let us now turn our attention to the structural evolution of 
the 2D-GCM ground state collective wave functions 
$G^{\sigma=1} (\vec{\beta})$  defined in Eq.(\ref{cll-wfs-HW}). Those 
wave functions are depicted in Fig.~\ref{GSCWF-examplesBa} for the 
same representative set of Ba isotopes as in Fig.~\ref{MFPES-examplesBa}.
A similar analysis has been performed for Xe, Ce and Nd nuclei.

From Fig.~\ref{GSCWF-examplesBa}, one realizes that for 
$^{118,122,126}$Ba (region $60 \le N \le 72$) the  collective wave 
function $G^{\sigma=1} (\vec{\beta})$ has a two-dimensional Gaussian 
shape with principal axes aligned along the directions A and B, which 
are tilted with respect to both the quadrupole and hexadecapole axes. 
The widths of the Gaussian are not the same, with the one along the A 
direction larger than the one corresponding to the B direction. As 
the neutron number increases, for $^{130,134,138}$Ba (region $74 \le N \le 
82$), the collective wave function keeps its Gaussian shape, but this 
time the principal axes are  aligned parallel to the $\beta_{2}$  and 
$\beta_{4}$ axes. Once more, for $^{142,146,152,158,162,168}$Ba (region 
$84 \le N \le 112$) the 2D-GCM Gaussian aligns along the tilted 
direction A. Finally, in the case of $^{174,178,182}$Ba (region $114 
\le N \le 126$) those wave functions are aligned parallel to the 
$\beta_{2}$ axis. The shape of the ground state collective wave function
$G^{\sigma=1} (\vec{\beta})$
closely follows the evolution of the corresponding MFPES around the ground
state minimum.

From a beyond-mean-field perspective, the previous results indicate 
that the quadrupole and hexadecapole degrees  are dynamically 
interwoven  in the ground states of Ba isotopes with $60 \le N \le 72$ 
and $84 \le N \le 112$, while they are decoupled for $74 \le N \le 82$ 
and $114 \le N \le 126$. Our analysis of the $G^{\sigma=1} 
(\vec{\beta})$ strengths in Xe, Ba, Ce and Nd isotopes corroborates the 
HFB prediction of two transitions from a 
$(\beta_{2},\beta_{4})$-coupled to a decoupled regime. Similar to the 
results obtained at the HFB level, the first  transition takes place 
for $N=70, 74, 76$ and $78$ in Xe, Ba, Ce and Nd isotopes and the 
second one occurs for $N=114$ in Xe, Ba and  Ce and for $N=118$ in Nd 
isotopes.

\begin{figure*}
\includegraphics[width=1.\textwidth]{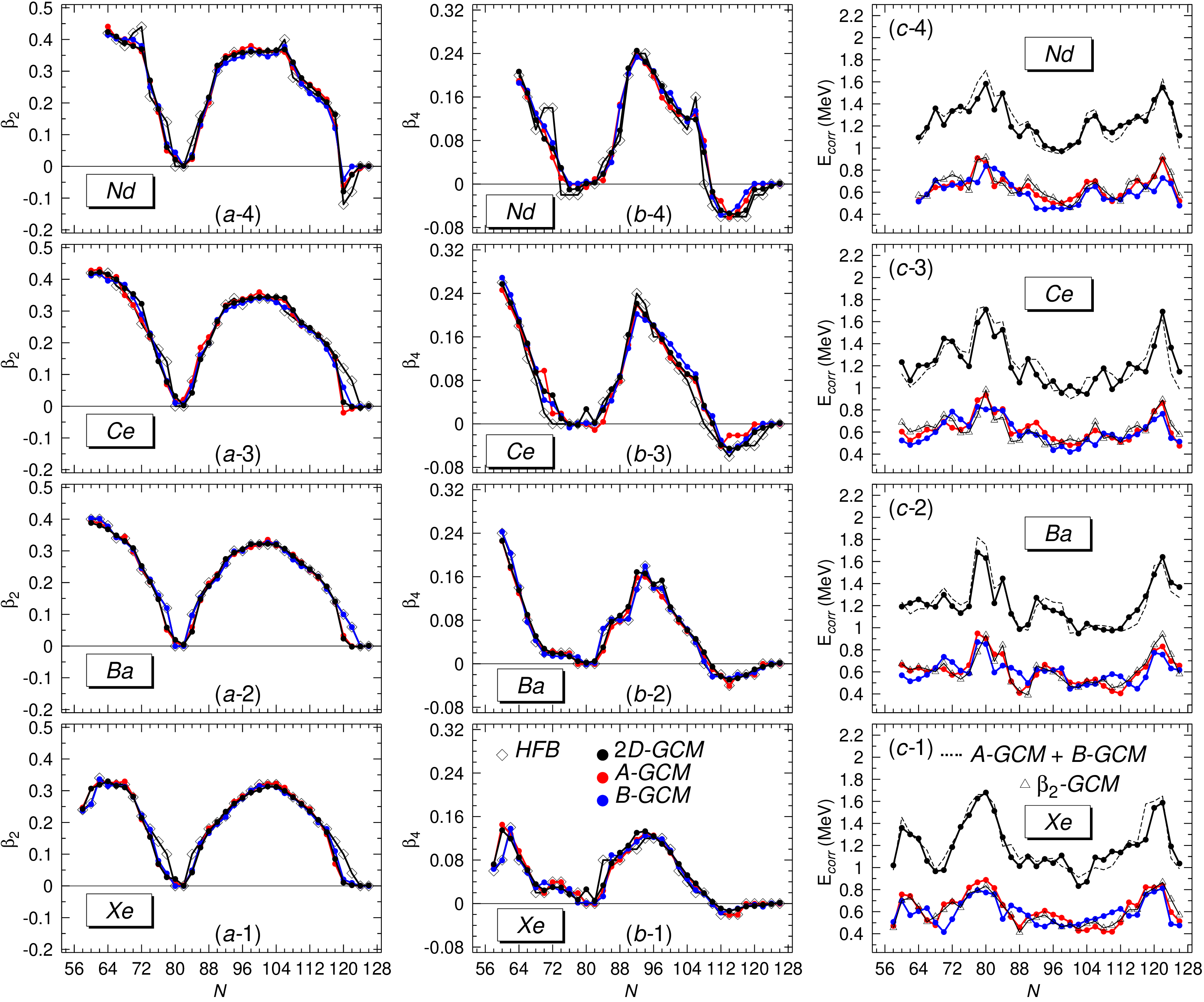}
\caption{(Color online)  2D-GCM ground state quadrupole $\beta_{2}$ 
[panels (a1)-(a4)] and hexadecapole $\beta_{4}$ [panels (b1)-(b4)] 
deformation parameters for $^{112-180}$Xe, $^{116-182}$Ba, $^{118-184}$Ce
 and $^{124-186}$Nd. The HFB ground state quadrupole 
and hexadecapole  deformations (Fig.\ref{HFB-b2b4-gs-def})
are included in panels (a1)-(a4) and (b1)-(b4) for comparison.
 The 2D-GCM correlation energies 
obtained for those nuclei are depicted in panels (c1)-(c4). 
Quadrupole and hexadecapole deformations as well as correlation 
energies obtained in 1D A-GCM and B-GCM calculations 
are included in panels (a1)-(a4), (b1)-(b4) and (c1)-(c4), respectively. 
The dotted lines in panels (c1)-(c4) 
correspond to the sum of A-GCM and B-GCM correlation energies. The quadrupole
correlation energies obtained in 1D $\beta_{2}$-GCM calculations are also
depicted in panels (c1)-(c4). Results are obtained with the Gogny-D1S EDF. 
For more details, see the main text.
}
\label{summary-2DGDM_GS} 
\end{figure*}

The dynamical ground state quadrupole and
hexadecapole deformations are shown in panels (a1)-(a4) and (b1)-(b4) of Fig.~\ref{summary-2DGDM_GS} along with
the HFB results, depicted for comparison. 
As can be seen from the panels, the inclusion of the quantum fluctuations in the 
$(\beta_{2},\beta_{4})$ degrees of freedom results in smoother evolution patterns of the deformations 
$\beta_{2,2D-GCM}^{\sigma=1}$ and   $\beta_{4,2D-GCM}^{\sigma=1}$, as compared with the 
mean-field values. The kinks observed at the mean-field level are smoothed out
when the quantum fluctuations are included. It is also remarkable to mention
that away from transitional regions the mean-field results for the quadrupole
and hexadecapole deformations follow closely the ones obtained with the
more sophisticated 2D-GCM calculation. 

Exception made of Xe isotopes,  for which there is 
an initial increase up to $0.33$ in $^{118}$Xe, the 
ground state quadrupole deformations in panels (a1)-(a4)
display an overall decreasing trend, reaching 
$\beta_{2,2D-GCM}^{\sigma=1}=0$ for $N=82$. Beyond $N=82$, those
deformations increase, reaching their largest values 
($\beta_{2,2D-GCM}^{\sigma=1}= 0.30-0.34$) around 
$N=102$ with a softened  kink 
($\beta_{2,2D-GCM}^{\sigma=1}= 0.37$) for $^{166}$Nd. For 
larger neutron numbers the $\beta_{2,2D-GCM}^{\sigma=1}$
values decrease, reaching $\beta_{2,2D-GCM}^{\sigma=1}=0$ 
for $^{176-180}$Xe, $^{178-182}$Ba, $^{180-184}$Ce 
and $^{184,186}$Nd. Note that, similar to the HFB 
results (see, Fig.~\ref{HFB-b2b4-gs-def}), a
transition to weakly oblate  
($\beta_{2,2D-GCM}^{\sigma=1}=-0.08$ and $-0.03$) states is 
predicted  for 
$^{180,182}$Nd.

%######################################################################
% figure 
%######################################################################

The pattern of the $\beta_{4,2D-GCM}^{\sigma=1}$ deformations in panels 
(b1)-(b4) confirms the (dynamical) interrelation with the 
$\beta_{2,2D-GCM}^{\sigma=1}$ deformations, i.e., large hexadecapole 
$\beta_{4,2D-GCM}^{\sigma=1}$ values are reached precisely in the 
regions corresponding to large quadrupole $\beta_{2,2D-GCM}^{\sigma=1}$ 
values \cite{inakura-2}. It is also apparent from the panels that 
dynamical hexadecapole deformations play a role in the ground states of 
most of the studied nuclei.

Exception made of the initial increase up to $0.13$ in $^{114}$Xe, the 
hexadecapole deformations display an overall decrease, reaching 
$\beta_{2,2D-GCM}^{\sigma=1}=0$ for $N=82$. As can be seen from panel 
(b4), the square-like shapes, predicted within the HFB approach for Nd 
isotopes below $N=82$ \cite{polar-gap-model} (see, 
Fig.~\ref{HFB-b2b4-gs-def}),  survive the effects of 2D 
quadrupole-hexadecapole configuration mixing in   $^{136,138}$Nd  
$(\beta_{4} \approx -0.01)$.

The $\beta_{4,2D-GCM}^{\sigma=1}$ parameters increase beyond $N=82$, 
reaching their largest  values $\beta_{4,2D-GCM}^{\sigma=1}=0.13, 0.16, 
0.22$ and $0.25$  in $^{148}$Xe, $^{148}$Ba, $^{150}$Ce and $^{152}$Nd. 
For larger neutron numbers, the ground state hexadecapole deformations 
decrease. For the studied Xe, Ba, Ce and Nd isotopic chains, our 2D-GCM 
calculations confirm the dynamical stability of a region, just below 
$N=126$ \cite{polar-gap-model}, with negative hexadecapole deformations 
within the range $-0.05 \le \beta_{4,2D-GCM}^{\sigma=1}=0 \le -0.01$. 
At the 2D-GCM level the nuclei in this region are $^{166-170}$Xe, 
$^{168-176}$Ba, $^{170-180}$Ce and $^{170-182}$Nd.

Given the dynamical relevance of the directions A and B, we have also 
performed 1D-GCM calculations along those directions. Let us stress 
that the directions A and B are characterized by the linear dependencies  
$\beta_{4,A}= a_{A} \beta_{2} + b_{A}$ and $\beta_{4,B}= -(1/a_{A}) 
\beta_{2} + b_{B}$. For each nucleus, we have fitted the parameters 
$a_{A}$, $b_{A}$ and $b_{B}$ using configurations around the absolute 
minimum of the corresponding MFPES 
\cite{large-scale-b4,Rayner-Robledo-b2-b4-RaThUPu,Rayner-Robledo-b2-b4-YbHfWOs}. 
In the A-GCM and B-GCM calculations, we have resorted to a 
$\beta_{2}$-mesh with $\beta_{2} \in [-0.80,0.92]$ and the step size 
$\delta \beta_{2}= 0.04$. The same $\beta_{2}$-mesh and step size have 
also been employed in 1D-GCM calculations with the quadrupole 
deformation $\beta_{2}$ as single generating coordinate 
($\beta_{2}$-GCM).

The quadrupole and hexadecapole deformation parameters obtained in the 
A-GCM and B-GCM calculations are also included in panels (a1)-(a4) and 
(b1)-(b4) of Fig.~\ref{summary-2DGDM_GS}. They agree rather well with 
the 2D-GCM quadrupole $\beta_{2,2D-GCM}^{\sigma=1}$ and hexadecapole 
$\beta_{4,2D-GCM}^{\sigma=1}$ deformations. Though the corresponding 
values are not included in panels (a1)-(a4) and (b1)-(b4), the same 
holds true for the $\beta_{2}$-GCM quadrupole and hexadecapole 
deformations. This is not surprising as the A-GCM and $\beta_{2}$-GCM
calculations explore nearly the same 
configurations around the ground state minimum.

%######################################################################
% figure 
%######################################################################

In panels (c1)-(c4) of Fig.~\ref{summary-2DGDM_GS}, we have plotted the 
2D-GCM 
correlation energies $E_{Corr,2D-GCM}$, defined as the difference 
\begin{equation}
E_{Corr,2D-GCM} = E_{HFB,gs}-E_{2D-GCM,gs}
\end{equation}
between the HFB $E_{HFB,gs}$ and 2D-GCM $E_{2D-GCM,gs}$
ground state energies. For the 
Xe, Ba, Ce and Nd chains, we have obtained values within the 
range $0.83~MeV \le E_{Corr,2D-GCM} \le 1.71~MeV$. As the range of variation of 
the energies $E_{Corr,2D-GCM}$ ($0.88~MeV$) is comparable with the rms deviation for 
the binding energies in Gogny mass tables 
\cite{gogny-d1mstar,gogny-d1m} we consider that $(\beta_{2},\beta_{4})$-fluctuations 
should be included in future fitting protocols of Gogny-EDF
\cite{Rayner-Robledo-b2-b4-RaThUPu,Rayner-Robledo-b2-b4-YbHfWOs}.

In panels (c1)-(c4), we have also included the 
A-GCM, B-GCM and $\beta_{2}$-GCM correlation energies. 
All those 1D-GCM correlation energies are rather similar.
First, the A-GCM and $\beta_{2}$-GCM energies are similar 
since the line A essentially represents the direction of quadrupole 
fluctuations \cite{large-scale-b4}. Second, the A-GCM and B-GCM energies are similar in magnitude. This
is a bit surprising result because the correlation energy can be estimated as the square root
of the ratio between the curvature of the MFPES and the collective inertia. It turns out that, in general,
the curvature along the B direction is larger than the one along the A direction, implying
that the same has to hold true for the collective inertias in order to 
have a roughly constant ratio \cite{Rayner-Robledo-b2-b4-RaThUPu,Rayner-Robledo-b2-b4-YbHfWOs}.

 Moreover, 
as can be seen from the panels, the 
sum of the A-GCM (or $\beta_{2}$-GCM) and B-GCM correlation energies accounts 
reasonably well for the 2D-GCM correlation energies $E_{Corr,2D-GCM}$. Within this
context, A-GCM and B-GCM calculations provide a cheaper 
alternative to roughly determine 2D-GCM ground state correlation energies
in mass table calculations. It has to be taken into account that the 
determination of the A direction can be carried out in a 1D HFB calculation
as a function of $\beta_{2}$, being the B direction determined by the
imposition to be orthogonal to the A one. 

We have obtained $\beta_{2}$-GCM correlation energies within the range 
$0.41~MeV \le E_{Corr,\beta_{2}-GCM} \le 0.98~MeV$. Thus, in going from 
the $\beta_{2}$-GCM to the full $(\beta_{2},\beta_{4})$-GCM 
calculations, one obtains an additional correlation energy gain 
($0.42~MeV \le \delta_{GCM} \le 0.73~MeV$) that compares well with the 
quadrupole correlation energy $E_{Corr,\beta_{2}-GCM}$ itself. This 
reinforces the conclusions extracted from previous calculations for Sm, 
Gd, actinides and rare earth elements regarding the key role of 
hexadecapole deformation in the ground state dynamics as well as the 
slow convergence of the nuclear correlation energies with respect to 
the multipole order of the associated shape deformation parameters 
\cite{large-scale-b4,Rayner-Robledo-b2-b4-RaThUPu,Rayner-Robledo-b2-b4-YbHfWOs}.

% ----------------------------------------------------------------------
%
% ----------------------------------------------------------------------
\subsubsection{First excited states} 
\label{sec-exc-states}
 
\begin{figure*}
\includegraphics[width=1.\textwidth]{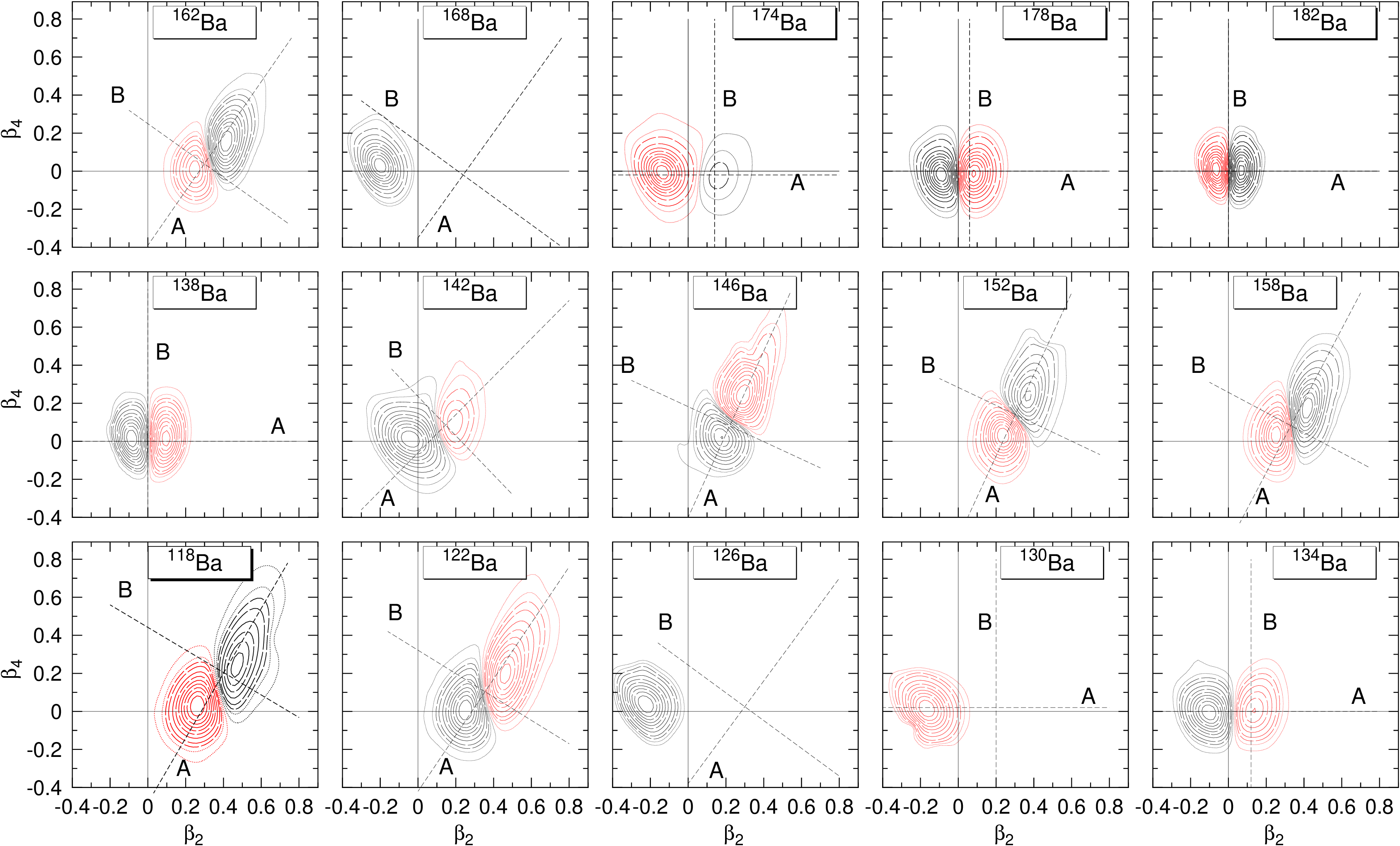}
\caption{(Color online) The same as in Fig.~\ref{GSCWF-examplesBa} but for the first excited 
states of a selected set of Ba isotopes.
Red
contour lines correspond to negative values of the collective wave functions. 
The transition region between red and black contours signals the existence
of a node in 
the collective wave function.}
\label{FEXCCWF-examplesBa} 
\end{figure*}

In Fig.~\ref{FEXCCWF-examplesBa}, we have depicted 
the collective wave 
functions $G^{\sigma=2} (\vec{\beta})$ Eq.(\ref{cll-wfs-HW})
corresponding to the first excited states for the 
same selected set of Ba isotopes as in 
Fig.~\ref{MFPES-examplesBa}.

%######################################################################
% figure 
%######################################################################

For $^{118,122}$Ba, the amplitudes $G^{\sigma=2} (\vec{\beta})$ are typical 
examples of the results obtained for 
$60 \le N \le 66$
isotopes. They correspond to a phonon aligned along 
the direction A. In those A-aligned phonons the quadrupole 
and hexadecapole degrees are interwoven. On the other hand, Ba isotopes
with $68 \le N \le 82$ belong to a $(\beta_{2},\beta_{4})$-decoupled 
regime with collective strengths aligned parallel
to the $\beta_{2}$-axis. First, in the region
$68 \le N \le 76$, nuclei such as $^{126,130}$Ba
exhibit collective 
amplitudes $G^{\sigma=2} (\vec{\beta})$ 
mainly
concentrated on the oblate side. Note that the 
ground state collective wave functions in 
$^{126,130}$Ba concentrate on the prolate 
side (see, Fig.~\ref{GSCWF-examplesBa}). Given 
the more pronounced shape coexistence 
observed in these isotopes 
[see, panel (a) of Fig.~\ref{SCoexistence}], it is not surprising
that their first excited states correspond 
to (oblate) shape isomers. On the other hand, for 
$^{132}$Ba (not shown in the figure) the 
collective strength $G^{\sigma=2} (\vec{\beta})$  
displays a prolate component
associated with the emergence 
of pure one-phonon quadrupole vibrations
in nuclei such as $^{134,138}$Ba ($78 \le N \le 82$).

For $^{142,146,152,158,162}$Ba (region $84 \le N \le 110$), the 
first excited states correspond once again to A-aligned phonons 
with the quadrupole and hexadecapole degrees interwoven.
The
alignment of the  
$G^{\sigma=2} (\vec{\beta})$ amplitudes parallel
to the $\beta_{2}$-axis signals the decoupling of the 
quadrupole and hexadecapole degrees for 
$112 \le N \le 126$
isotopes. First, in the region
$112 \le N \le 118$, a nucleus such as $^{168}$Ba
exhibits a collective 
amplitude $G^{\sigma=2} (\vec{\beta})$ 
corresponding to an oblate shape isomer, which is
a consequence of the pronounced competition between 
prolate and oblate configurations in the region
[see, panel (b) of Fig.~\ref{SCoexistence}]. On the other
hand, for 
$^{174}$Ba (similar to the case of $^{132}$Ba) the 
collective strength $G^{\sigma=2} (\vec{\beta})$  
 displays a prolate component
associated with the emergence of pure two-phonon 
quadrupole vibrations
in nuclei such as $^{178,182}$Ba ($120 \le N \le 126$)

From the analysis of the structural evolution of the amplitudes 
$G^{\sigma=2} (\vec{\beta})$, one realizes that not only for the ground 
but also for the first excited states two transitions occur from a 
$(\beta_{2},\beta_{4})$-coupled to an decoupled regime. The first 
(second) transition takes place for $N=64, 68, 74$ and $74$ ($N=110, 
112, 114$ and $114$) in Xe, Ba, Ce and Nd isotopes. 

%######################################################################
% figure 
%######################################################################
\begin{figure*}
\includegraphics[width=1.\textwidth]{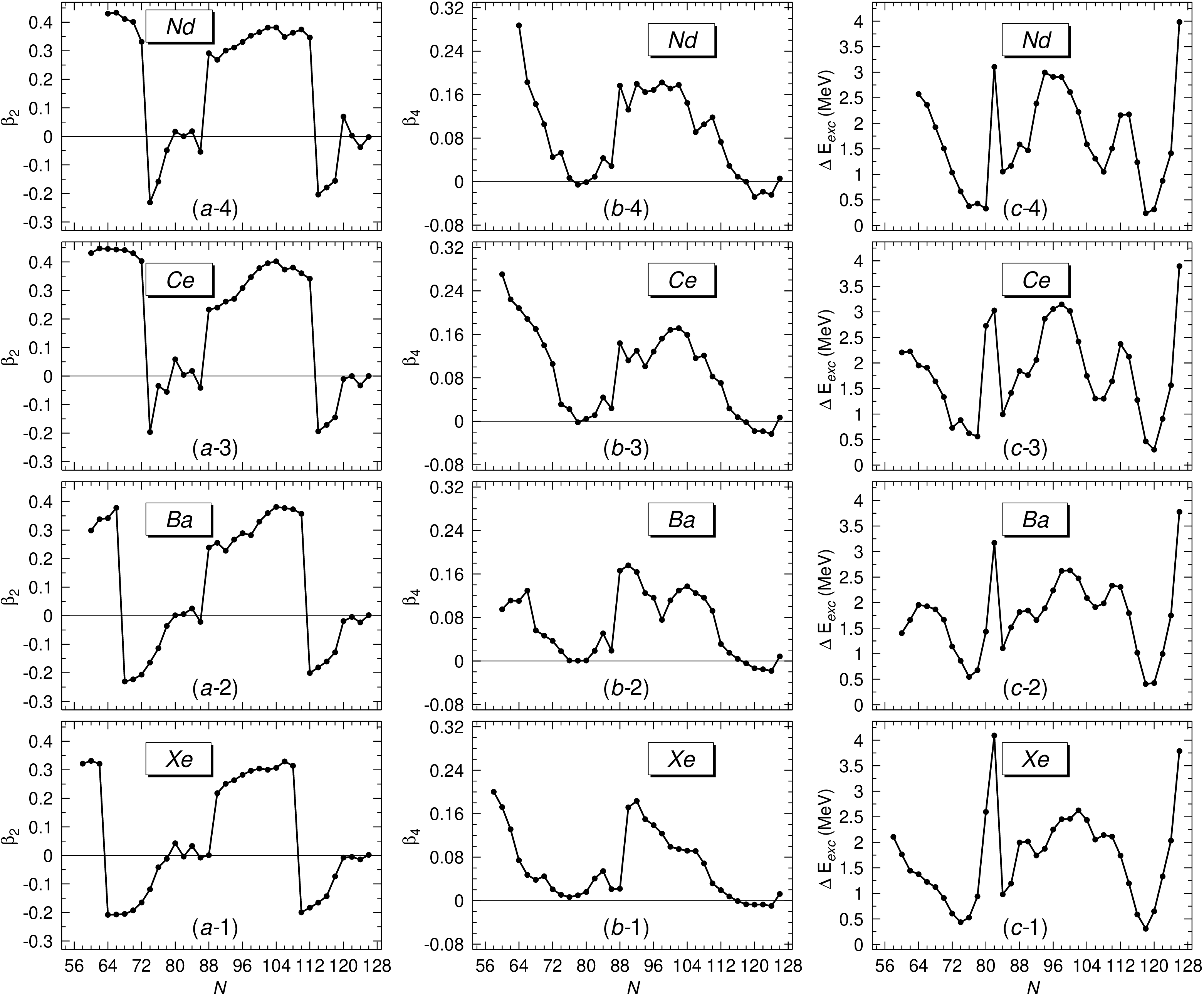}
\caption{2D-GCM quadrupole $\beta_{2}$ [panels (a1)-(a4)] and 
hexadecapole $\beta_{4}$ [panels (b1)-(b4)] deformation parameters 
corresponding to the first excited states in
$^{112-180}$Xe, $^{116-182}$Ba, $^{118-184}$Ce
 and $^{124-186}$Nd. The excitation 
energies $\Delta E^{ \sigma=2}$ corresponding to those states are 
depicted in panels (c1)-(c4). Results are obtained with the 
Gogny-D1S EDF. For more details, see the main text. 
}
\label{summary-2DGDM_FEXC} 
\end{figure*}

The quadrupole $\beta_{2,2D-GCM}^{\sigma=2}$ deformations 
are plotted in panels (a1)-(a4) of Fig.~\ref{summary-2DGDM_FEXC}.
As can be seen from the panels, the nuclei 
$^{112-116}$Xe, $^{116-122}$Ba, $^{118-130}$Ce 
and $^{124-132}$Nd exhibit prolate $(0.13 \le \beta_{2} \le 0.45)$
 first excited states while,   
oblate $(-0.23  \le \beta_{2,2D-GCM}^{\sigma=2} \le -0.03)$
shapes are predicted for $^{118-130}$Xe, $^{124-134}$Ba, $^{132-136}$Ce
and $^{134-138}$Nd. The $\beta_{2,2D-GCM}^{\sigma=2}$ 
parameters reach
values close and/or equal to  
zero around $N=82$. This is followed by a region 
characterized by sizeable  prolate deformations, reaching their largest 
values $\beta_{2,2D-GCM}^{\sigma=2}=0.33, 0.38, 0.40$ and $0.38$
for $^{160}$Xe, $^{160}$Ba, $^{162}$Ba and $^{162}$Nd. 
A transition to oblate 
$(-0.20  \le \beta_{2,2D-GCM}^{\sigma=2} \le -0.01)$
first excited states is predicted in heavier isotopes as the neutron magic number 
$N=126$ is approached.

From panels (b1)-(b5) of Fig.~\ref{summary-2DGDM_FEXC}, one realizes
that dynamical hexadecapole $\beta_{4,2D-GCM}^{\sigma=2}$ deformations
play a role 
not only in the ground
(see, Fig.~\ref{summary-2DGDM_GS}) but also in the first excited states of the 
considered chains. Exception made of Ba isotopes,  for which there is 
an initial increase up to $0.13$ in $^{122}$Ba, those 
deformations exhibit a decreasing trend, reaching 
$\beta_{4,2D-GCM}^{\sigma=2}=0.01$ for $^{130}$Xe
and $\beta_{4,2D-GCM}^{\sigma=2}=0$ for 
$^{132-136}$Ba, $^{136}$Ce and $^{138}$Nd. 
Another distinctive feature in panels (b1)-(b4) is the presence of 
a region of diamond-like first excited states characterized by sizeable 
$\beta_{4,2D-GCM}^{\sigma=2}>0$
values. Note, that first excited states in this region also 
display large quadrupole deformations [panels (a1)-(a5)]
\cite{inakura-2}. For Xe, Ba, Ce and Nd isotopes, 
the largest 
hexadecapole deformation ($\beta_{4,2D-GCM}^{\sigma=2}=0.18, 0.18, 0.17$ and $0.18$)
in the region corresponds to $^{146}$Xe, $^{146}$Ba, $^{160}$Ce and $^{158}$Nd.
Furthermore, just below 
$N=126$  \cite{polar-gap-model}, our 2D-GCM calculations 
predict square-like first excited states 
in $^{172-178}$Xe, $^{176-180}$Ba, $^{178-182}$Ce
and $^{180-184}$Nd with negative 
hexadecapole deformations within 
the range $ -0.03 \le \beta_{4,2D-GCM}^{\sigma=2} \le -0.01$.

Finally, the excitation energies $\Delta E^{ \sigma=2}$ of the $K=0$ 
excited states are depicted in panels (c1)-(c4) of 
Fig.~\ref{summary-2DGDM_FEXC}. As functions of the neutron number they 
display two pronounced minima. For the first (second) minimum in 
$^{128}$Xe, $^{132}$Ba, $^{136}$Ce and $^{140}$Nd ($^{172}$Xe, 
$^{174}$Ba, $^{178}$Ce and $^{178}$Nd) those energies reach the values 
$\Delta E^{ \sigma=2}=$ 0.43, 0.54, 0.56 and 0.33~MeV ($\Delta E^{ 
\sigma=2}=$ 0.31, 0.41, 0.30  and 0.24~MeV). It is precisely the 
enhanced shape coexistence, as one moves towards the $N=82$ and $126$ 
shell closures, what leads to such low values of the excitation energies $\Delta 
E^{ \sigma=2}$. The energies $\Delta E^{ \sigma=2}$ also exhibit local 
minima around $N=108$. As already discussed in 
Sec.~\ref{static-HFB-TeXeBaCeNd} [see, panel (b) of 
Fig.~\ref{SCoexistence}], these minima reflect the subtle balance 
between competing configurations around this neutron number. Note that, 
in agreement with the HFB results, these local  minima become deeper 
with increasing $Z$ values. Furthermore, the pronounced maxima observed 
in the energies $\Delta E^{ \sigma=2}$ obtained for $^{136}$Xe, 
$^{138}$Ba, $^{140}$Ce and $^{142}$Nd as well as $^{180}$Xe, 
$^{182}$Ba, $^{184}$Ce and $^{186}$Nd represent the strong signatures 
of the $N=82$ and $126$ neutron shell closures, respectively.

%----------------------------------------------------------------------
%
%                                                  C o n c l u s i o n s
%
% ----------------------------------------------------------------------

\section{Conclusions}
\label{conclusions}

The coupling between the axial quadrupole and hexadecapole moments has
been analyzed from a dynamical perspective in the 
Xe, Ce, Ba and Nd isotopic chains. The range of neutron numbers considered is large enough as to cover the 
neutron magic numbers N=82 and N=126. This range is required to study 
the transition from positive to negative hexadecapole deformations. Two
regions of strong coupling between the quadrupole and hexadecapole collective degrees
of freedom are identified. Another two regions, close to neutron magic numbers,
where the quadrupole and hexadecapole degrees are decoupled are also analyzed. In 
regions of strong coupling a full-fledged 2D-GCM dynamical treatment
is required for a detailed treatment of the 
problem. However, by identifying orthogonal 
directions in the 
$(\beta_{2},\beta_{4})$-plane, along the 2D approximate 
parabolic potential energy surface 
around the ground state configuration, it is possible to obtain results
pretty close to the 2D-GCM ones by carrying out separate 1D GCM calculations
along the two perpendicular directions. The same holds true in the
regions where quadrupole and hexadecapole are decoupled, but in this 
case the identification of the orthogonal directions is not required. 

Quantum fluctuations soften some kinks observed in 
the ground state's $\beta_{2}$ and $\beta_{4}$ values in transitional 
regions. Away from those transitional regions the 2D-GCM and HFB
results are comparable. Another interesting result is the large value
of the zero-point energy associated to the hexadecapole degree of freedom.
It is comparable to the one of the quadrupole degree of freedom and both
together amount to roughly 1.5 MeV. This large zero-point energy has to
be taken into account for a proper definition of nuclear mass tables as
typical rms of binding energies is half the above value.

The properties
of the first excited state, in most of the cases, a one-phonon excitation along the 
lowest curvature
direction in the 2D MFPES around the ground state, are also
discussed. It turns out that in some regions of small $\beta_{2}$ 
deformation, shape coexistence between the prolate ground state and the oblate excited
minimum is observed. In those cases, the excitation energy of the first
excited state is pretty low, a result that can be considered as an experimental
fingerprint of shape coexistence. Finally, the relevant role played by 
the so-called second intruder configuration in  determining the large deformation parameters
obtained for the ground state mean field configuration  of several isotopes in the region  is analyzed in a couple of relevant
examples.

%------------------------------------------------------------------------
\begin{acknowledgments}
%-----------------------------------------------------------------------
The work of R. Rodr\'{\i}guez-Guzm\'an and K. Ziyatkhan is funded by 
Nazarbayev University under the Faculty Development Competitive 
Research Grants Program (FDCRGP) for 2025-2027, Grant 040225FD4712. The 
work of LMR is supported by Spanish Agencia Estatal de Investigacion 
(AEI) of the Ministry of Science and Innovation under Grant No. 
PID2024-159559NB-C21. Both R.R.-G. and K.Z. also acknowledge the 
support provided by the IT Department for performing calculations on 
the HPC clusters Shabyt and Irgetas at Nazarbayev University.
\end{acknowledgments}

\end{document}